\documentstyle[12pt]{article}
\begin{document}
\huge
\begin{center}
\Large{{\bf THE STRING PHASES OF

 HAWKING RADIATION, DE SITTER STAGE 

AND DE BROGLIE TYPE DUALITY}}\\ 
\vspace{1cm}
{\bf Marina RAMON MEDRANO${}^1$ , 

Norma G. SANCHEZ${}^2$}
\vspace{1cm}

[1] Departamento de F\'{\i}sica Te\'orica, Facultad de Ciencias
F\'{\i}sicas, Universidad Complutense, E-28040, Madrid, Spain.

email: mrm@fis.ucm.es

[2] Observatoire de Paris, LERMA (Laboratoire Associ\'e au CNRS 
UMR 8112, Observatoire de Paris, Univ. Paris VI et Ecole Normale Sup\'erieure), 
61 Avenue de l'Observatoire, 75014 Paris,France.

email: Norma.Sanchez@obspm.fr

\vskip 1cm
\begin{abstract}
We explicitely describe the last stages of black hole evaporation in the context of string theory : the 
combined study of Quantum Field Theory (QFT) and String Theory (ST) in curved backgrounds allows us to go further in the 
understanding of quantum gravity effects. The string ``analogue model''(or thermo-dynamical approach) is a well suited 
framework for this purpose.The results also apply to another physically  relevant case: de Sitter background. 
Semiclassical (QFT)  and
quantum gravity (String) phases or regimes are properly determined (back reaction effects included).  
The Hawking-Gibbons temperature ${T_H}$ of the semiclassical regime becomes the intrinsic string temperature ${T_S}$ 
in the quantum gravity regime.
The spectrum of black hole evaporation is an incomplete gamma function of $(T_S - T_H)$: 
the early evaporation is thermal (Hawking radiation), while at the end the black hole undergoes a phase transition to a 
string state decaying (as string decay) into pure (non mixed) particle states. Remarquably, explicit dynamical 
computations show that  both gravity regimes: semiclassical (QFT) and quantum (string), are dual of each other, 
in the precise sense of the classical-quantum (de Broglie type) duality. 
\end{abstract}

\end{center}
\normalsize
\newpage

\newpage

\section{Introduction}
The combined study of Quantum Field Theory (QFT) and Quantum String Theory (QST) 
in curved space times - and in the framework of the string analogue model -
allows to go further in the understanding of quantum gravity effects Ref. [1,2,3,4,16,17]. 

\medskip
The string ``analogue model'' (or thermodynamical approach) is a suitable 
framework for this purpose.
Strings are considered as a collection of quantum fields $\phi_n$ which do not
interact among themselves but are coupled to the curved 
background, and whose masses are given by the degenerated string spectrum in 
this background. The 
higher mass spectrum is described by the density of string mass levels $\rho (m)$ in the space time
considered Ref. [3,4]. \\

In Black Hole backgrounds (BH), $\rho (m)$ is the same as in flat space time
and the string critical temperature is the usual (Hagedorn) temperature of strings
in flat space time Ref. [5]. In de Sitter (dS) as well as in Anti de Sitter (AdS)
backgrounds, (including the corresponding conformal invariant WZWN SL(2,R) models), $\rho (m)$ are different from flat 
space time Ref. [3,6,7,8,9,17]. There is a
critical (maximal) string temperature in de Sitter background, while in AdS
there is no finite maximal string temperature at all. Ref. [6,7,9] ( the
"maximal" temperature is formally infinite, as the partition function for a gas
of strings in AdS is defined for any external temperature Ref. [6]).

\medskip
The explicit results of quantum string dynamics in curved backgrounds developed in refs [3], [4],[16],[17] 
lead to a ${\cal R}$ ``dual'' (classical-quantum) transformation which, in particular, over a length $L$, simply  reads: 

\begin{equation} 
\tilde{L} = {\cal R} L = {\cal L}_{\cal R} \: L^{-1}                    
\end{equation}

\bigskip
\noindent
where ${\cal L}_{\cal R}$ has dimensions of $({\rm length})^2$. For physical processes, ${\cal R}$ maps 
classical length scales $L_{cl}$ into quantum string length 
scales $L_q$, and conversely: \\
\begin{equation} 
\tilde{L}_{cl} \equiv {\cal R} L_{cl} = {\cal L}_{\cal R} \: L^{-1}_{cl} = \: L_q  
\end{equation}
and
\begin{equation} 
\tilde{L}_q  \equiv {\cal R} L_q = {\cal L}_{\cal R} \: L^{-1}_q = \: L_{cl}  
\end{equation}

\bigskip
${\cal L}_{\cal R}$ depends on the
dimensional parameters of the theory and on the string constant $\alpha^{\prime}$ (and on $\hbar$, $c$ as well). 
The ${\cal R}$ transform is not an assumed or {\it a priori}
imposed symmetry, but it is a transformation {\it revealed} from the QFT and QST explicit dynamical calculations
in curved backgrounds Ref. [1,2,3,4,16,17]. 


\noindent
$L_{cl}$ sets up a length scale for the semiclassical - QFT regime and
$L_q$ is the length scale that  
characterizes the string domain; $L_q$ depends on the dimensional
string constant $\alpha^{\prime}$ 
($\alpha^{\prime} = c^2/2 \pi T$, where $T$ is the string tension) and
on the specific background considered. \\  
The "vehicle" of the ${\cal R}$-transformation is the dimensional constant
$\alpha^{\prime}$ of  string theory (ST). The 
${\cal R}$-operation transforms the characteristic lengths of one
regime (QFT or QST) into the characteristic lengths of the other, and thus relate 
different physical regimes in the {\it same} curved background. \\

\section{Classical-Quantum Duality Relations}

The ${\cal R}$-relation above described {\it does not needs  a priori} any symmetry or compactified 
dimensions. It {\it does not} require the existence of any isometry in the curved 
background. Different types of relativistic quantum type operations $L
\longrightarrow L^{-1}$  
appear in string theory due to the existence of the dimensional string
constant  
$\alpha^{\prime}$,
linking different 
equivalent string theories (the most known is T-duality). refs [12], [13]. The duality here we are considering is of the type
classical-quantum (or wave-particle) duality relating classical/semiclassical 
and quantum behaviours or regimes, during the {\it same physical process}. For example, the process of black hole 
evaporation or that of cosmological evolution, refs [16],[17].

\medskip
 Here we present evidence 
for classical-quantum ${\cal R}$-dual relations, without requiring any symmetry, isommetry or compactified dimensions. This evidence 
is supported by explicit dynamical computations of QFT and string dynamics in curved backgrounds and we illustrate it with two relevant 
examples: black holes and de Sitter space-time. (AdS states can be included, an explicit discussion is given in  ref [17]).

  This ${\cal R}$- transformation
being of the type  of a classical-quantum 
duality relationship, we do not attach a priori any symmetry to it.
Is known the role played by (global or asymptotic)  isometries in QFT
(and strings) on curved  backgrounds, for instance for identifying
the particle oscillatory  modes, the presence of event horizons,
global or asymptotic thermality, etc;  however such symmetries are
{\it not} responsible of the classical-quantum duality.

An enormous amount of work was 
devoted recently to the holographic AdS/CFT correspondence refs [14],
[15]. Our work does not make use of the AdS/CFT conjecture. 

\medskip
QFT in curved backgrounds with event horizons posseses an intrinsic
(Hawking-Gibbons) temperature $T_H$ 
Ref. [1,10,11], which can be expressed, in general, as a function $T$
of $L_{cl}$ (and of the constants $\hbar$, $c$ and $k_B$)  

\begin{equation}
T_H = T (L_{cl})                                                
\end{equation}

\bigskip
Quantum strings in Minkowski space time have an intrinsic (Hagedorn) temperature. Quantum strings in 
curved backgrounds have also an intrinsic string temperature $T_S$ Ref. [2,3,6], which depends on 
$L_q$ 

\begin{equation}
T_S = T (L_q)                                                   
\end{equation}

\bigskip
\noindent
Explicit calculations for de Sitter (dS) and Schwarzschild Black Hole (BH) show
that $T$ is formally the same function for both QFT and QST temperatures 
([Eq. (4)] and [Eq. (5)]), Ref. [3,4]. \\

\medskip
It is worth to point out that space times without event horizons -- such as anti de Sitter (AdS) space
-- have neither $T_H$ nor $T_S$ temperatures, independently of how much quantum matter is present. In fact, in pure AdS space, $T_H$ is zero and $T_S$ is formally 
infinite Ref. [6,9]. \\

\medskip
Applying the ${\cal R}$ operation ([Eq. (2)] to [Eq. (4)]) and [Eq. (5)], we read \\

\begin{equation}
\tilde{T}_H = T_S \: \: \: \: , \tilde{T}_S = T_H                      
\end{equation}

\bigskip
That is, $T_H$ and $T_S$ are ${\cal R}$-mapped one into the other. From the above
equations, we can also write 

\begin{equation}
\tilde{T}_H \: \tilde{T}_S  = T_S T_H                            
\end{equation}

\bigskip
\noindent
which is a weaker ${\cal R}$-relation between $T_H$ and $T_S$. \\

\medskip
Let us analyse the mass domains in the corresponding QFT and QS
regimes in a curved space time. They 
will be limited by the corresponding mass scales $M_H$ and
$M_{QS}$. $M_H$ depends on the classical  length $L_{cl}$ \\

\begin{equation}
M_H = M (L_{cl})                                                 
\end{equation}

\bigskip
\noindent
and $M_{QS}$ depends on the quantum length $L_q$, through the same formal
relation. Refs [16], [17] \\

\begin{equation}
M_{QS} = M (L_q)                                                 
\end{equation}
 
\bigskip
Under the ${\cal R}$ operation [Eq. (2)], the scales of mass satisfy: \\

\begin{equation}
\tilde{M}_H = M_{QS} \: \: \: \: , \tilde{M}_{QS} = M_H                 
\end{equation}

\bigskip
On the other hand, if $m_{QFT}$ is the mass of a test particle in the QFT regime and $m_S$ the mass of
a particle state in the quantum string spectrum, the mapping of the QFT mass domain ${\cal D}$ onto the
QS mass domain -- and viceversa -- reads \\

\begin{equation}
{\cal R} \left( {\cal D} ( m_{QFT} , M_H ) \right) = {\cal D} (m_S , M_{QS} )       
\end{equation}

\bigskip
(See refs [16], [17] for more details). Summarizing: The QFT regime, characterized by ($L_{cl}$, $T_H$ and $M_H$),and 
the QS regime -- 
characterized by ($L_q$ , $T_S$ and $M_{QS}$) -- , both in a curved space background, are mapped one into
another under the ${\cal R}$-transform.  The set ($L_q$ , $T_S$, $M_{QS}$) is the quantum string dual of the 
classical/semiclassical QFT set ($L_{cl}$, $T_H$, $M_H$). Quantum gravity regime and classical/ semiclassical gravity 
regime are ${\cal R}$ dual of each other.\\

\medskip
We illustrate this duality relation and its meaning with two relevant examples: de Sitter (dS) and Black Hole (BH) space times. 
\section{QFT and QST in de Sitter space time}
The classical $L_{cl}$, or horizon radius, is 

\begin{equation}
L_{cl} = c H^{-1}                                               
\end{equation}

\bigskip
\noindent
where $H$ is the Hubble constant. The mass scale $M_H$ is such that $L_{cl}$ is its Compton wave 
length \\

\begin{equation}
M_H =  \frac{\hbar}{c L_{cl}}  = \: \frac{\hbar H}{c^2} \quad ,
\end{equation}

\bigskip
\noindent
and the QFT Hawking-Gibbons temperature is \\

\begin{equation}
T_H = \: \frac{\hbar}{2 \pi k_B c} \:  \kappa                   
\end{equation}

\bigskip
\noindent
here $\kappa$ is the surface gravity. For dS space time $\kappa = cH$, and $T_H$ reads \\
 
\begin{equation}
T_H = \: \frac{\hbar H}{2 \pi k_B} \: = \frac{\hbar c}{2 \pi k_B} \: \left( \frac{1}{L_{cl}} \right)
\end{equation}

\bigskip
On the other hand, canonical as well as semiclassical quantization of quantum strings in dS space time
Ref. [6,7,8,] lead to the existence of a maximum mass $m_{\max}\sim c(
\alpha 'H)^{-1}$  
for the quantum string (oscillating or
stable) particle spectrum. This maximal mass identifies the mass scale $M_{QS}$
 in the string regime: \\

\begin{equation}
M_{QS} \equiv m_{\max} \simeq c ( \alpha^{\prime} H )^{-1}                    
\end{equation}

\bigskip
The fact that there is a maximal mass $M_{QS}$ implies the existence of a (minimal) quantum string 
scale $L_q$, which is the corresponding Compton wave length \\

\begin{equation}
L_q = \: \frac{\alpha^{\prime} \hbar H}{c^2} \: = \: \frac{\hbar}{c M_{QS}}  \quad ,
\end{equation}

\bigskip
\noindent
and of a maximum (or critical) temperature $T_S$,  \\

\begin{equation}
T_S = \: \frac{\hbar c}{2 \pi k_B} \: \left( \frac{1}{L_q} \right) \: = \: \frac{c^3}{2 \pi k_B
\alpha^{\prime} H}                                                             
\end{equation}

\bigskip
This temperature is the string temperature emerging from the asymptotic (highly excited) mass spectrum of strings in 
de Sitter background. This temperature also appears to be the intrinsic temperature of de Sitter stage in its quantum 
gravity (string) regime, refs [3], [17]. (As is known, there is no de Sitter string conformal invariant background from 
the low effective string eqs, see refs [3], [17]).

$L_q$, $T_S$ and $M_{QS}$ depend on the string tension and on $H$,  while 
${\cal L}_{\cal R}$ [Eq. (3)] does only on $\alpha^{\prime}$:

\begin{equation}
{\cal L}_{\cal R} = \alpha^{\prime} \hbar c^{-1} \equiv L^2_S                  
\end{equation}

\medskip
\noindent
$L_S$ is a pure string scale. 

\medskip
The ${\cal R}$-transform maps the semiclassical/ QFT set 
($L_{cl}$, $T_H$ and $M_H$) into the quantum string set ($L_q$, $T_S$
and $M_{QS}$), satisfying the duality relations [Eqs. (2,6 and 10)]. 

\medskip

\bigskip
If back reaction effect of the quantum matter is considered,
${\cal R}$-relations between the QFT and QST regimes manifest as well, ref [3]. We
studied quantum string back reaction due to the higher excited modes in the 
framework of the string
analogue model. Two branches $R_{\pm}$ (i.e. $H_{\pm}$) of solutions for the 
scalar curvature show up ref. [3]:

\begin{description}
\item[(a)]
A high curvature solution $R_+$ with a maximal value $R_{\max} = (9 c^4 \pi^2 /4G) (6/(5 
\alpha^{\prime} c \hbar^3))^{1/2}$, entirely sustained by the strings. This branch corresponds to the 
string phase for the background, whose temperature is given by the {\it intrinsic string de Sitter temperature} 

\begin{equation}
T_S \equiv T_+ = \: \frac{c^3}{2 \pi k_B \alpha^{\prime} H_+} \qquad \qquad \qquad \qquad \cdot  
\end{equation}

\medskip

\item[(b)]

A low curvature solution $R_-$ whose leading term in ${R}/{R}_{\max}$ expansion is the
classical curvature. This branch corresponds to the semi-classical-QFT phase of the background, whose
temperature is given by the {\it intrinsic QFT de Sitter (Hawking-Gibbons) temperature} 

\begin{equation}
T_H \equiv T_- = \: \frac{\hbar H_-}{2 \pi k_B}                                            
\end{equation}

\end{description}
\bigskip
Then, the dual classical-quantum ${\cal R}$-relations 
manifest as well when {\it back reaction is included} : the stringy regime {\bf (a)} at 
the string temperature $T_S$ eq (21) and the semiclassical phase {\bf (b)} at the 
QFT Hawking-Gibbons temperature $T_H$ eq (22) are the quantum-classical $\mathcal{R}$-dual of each 
other. The branches ($R_+$, $T_+$) and ($R_-$, $T_-$) are the ${\cal R}$-transformed of each another.
\section{QFT and QST in Schwarzschild BH space time}
Here $L_{cl}$ is the BH radius $r_H$ ( $G$: gravitational Newton constant):

\begin{equation}
r_H = \: \left( \frac{16 \pi G M_H}{c^2 (D-2) A_{D-2}} \right)^{\frac{1}{D-3}} \qquad \qquad , \: 
\left( A_{D-2} \equiv \: \frac{2 \pi^{\frac{(D-1)}{2}}}{\Gamma \left( \frac{D-1}{2} \right)} \right) 
\end{equation}

\bigskip
\noindent
and $M_H$ is the BH mass. The QFT Hawking temperature is \\

\begin{equation}
T_H = \: \frac{\hbar c (D-3)}{4 \pi k_B} \: \left( \frac{1}{r_H} \right)               
\end{equation}

\bigskip
\noindent
($\kappa = (D-3) c^2 / 2 r_H$).In string theory, BH emission is described by an incomplete gamma function of ($T_S$ - $T_H$)Ref. [4]. In the QFT regime, the BH does emit thermal radiation at a temperature $T_H$ Ref. [10]. In 
the QST regime,
the BH has a high massive string emission, corresponding to the higher excited quantum string
states Ref. [4]. For open strings (in the asymptotically flat BH region), the thermodynamical behaviour of these
states is deduced from the string canonical partition function Ref. [4]. In BH backgrounds, the mass
spectrum of quantum string states coincides with the one in Minkowski space, and critical dimensions 
are the same as well, Ref. [5]. Therefore, the asymptotic string mass density of levels, in BH space 
times, reads $\rho (m) \sim \exp \{ b (\alpha^{\prime} c / \hbar )^{1/2} m \}$ , and quantum strings 
have an intrinsic temperature $T_S$ which is the same as in flat space time. The string canonical 
partition function is defined for Hawking temperatures $T_H$ satisfying the condition Ref. [4] \\

\begin{equation}
T_H  < T_S = \: \frac{\hbar c}{b k_B L_S}                            
\end{equation}

\bigskip
\noindent
i.e in string theory, $T_H$ has an upper limit given by the intrinsic or critical string temperature 
$T_S$. This limit implies the existence of a minimal BH radius $r_{\min}$, and a minimal BH mass
$M_{\min}$ 
\begin{equation}
 r_H > r_{\min} =  \: \frac{b (D-3)}{4 \pi} \: L_S 
\end{equation}
 and
\begin{equation}
M_H > M_{\min} =  \: \frac{c^2 (D-2) A_{D-2}}{16 \pi G} \: \left( \frac{b (D-3) L_S}{4 \pi} 
\right)^{D-3} 
\end{equation}
\noindent
Here $L_S$ is given by [Eq. (19)], and ${\cal L}_R$ [Eqs.(2,3)] is given by

$${\cal L}_R={b M_H \alpha^{\prime {3 \over 2}} \over 2 \pi } \sqrt{{\hbar
    \over c}}, \qquad D=4\,. $$ 

\bigskip
Therefore, the QS scales are $L_q = r_{\min}$ , $M_{QS} = M_{\min}$ ,
and the string temperature $T_S$. 
They depend on the type of strings and on the dimension through the
parameter $b$. In terms of  
$r_{\min}$, $T_S$ and $M_{\min}$ read \\
\begin{eqnarray}
T_S  & = & \: \frac{\hbar c (D-3)}{4 \pi k_B} \: \left( \frac{1}{r_{\min}} \right) 
\\ \cr
M_{\min} & = & \frac{c^2 (D-2) A_{D-2}}{16 \pi G} \: r^{D-3}_{\min}  
\end{eqnarray}
\noindent
The set ($r_H , M_H , T_H$) Eqs.(22)-(23) and the set ($r_{\min} , M_{\min} ,T_S$) Eqs.(24)-(26) are ${\cal R}$-transformed of each other: 
the set ($r_{\min} , M_{\min} ,T_S$) is the quantum dual of the semiclassical set ($r_H , M_H , T_H$). 
This is valid in all dimensions. \\

\medskip
The physical meaning (shown by the explicit results of Ref.4) behind these  classical-quantum ${\cal R}$-relations 
is the following. At the first stages of BH 
evaporation, emission is in the lighter particle masses at the Hawking temperature $T_H$, as described
by the semiclassical - QFT regime. As evaporation proceeds, temperature increases and high massive 
emission
corresponds to the higher excited string modes. At the later stages, for $T_H \to T_S$ (i.e $r_H \to
r_{\min}$ , $M_H \to M_{\min}$), the BH enters its QS regime. The ${\cal R}$ 
transformation allows to link
the early or semiclassical ($T_H \ll T_S$, i.e $r_H \gg r_{\min}$, $M_H \gg M_{\min}$) and the late or stringy ($T_H \to T_S$ , i.e 
$r_H \to r_{\min}$ , $M_H \to M_{\min}$) stages of BH evaporation. \\

\medskip
These classical-quantum ${\cal R}$-relations manifest as well if the {\it back reaction effect of higher massive string modes is included} Ref. [4]: The string back reaction solution ($r_+$, $M_+$, $T_+$) shows that the BH radius $r_+$ and
mass $M_+$ decrease and the BH temperature $T_+$ increases. Here $r_+$ is bounded from below (by
$r_{\min}$) and $T_+$ {\it does not blow up} ($T_S$ is the maximal value). The string back reaction effect is
finite and consistently describes both the QFT ($T_H \ll T_S$) and the QST ($T_H \to T_S$) regimes. It 
has the bounds: $(r,T,M)_{\min} < (r,T,M)_+ < (r,T,M)_H$. The two bounds are linked by the ${\cal R}$
transform, they are the classical-quantum dual of each other, ie the ${\cal R}$ transform links the early (QFT) and the 
last (QST) stages of evaporation.

\section{Conclusions}
From the above BH and dS studies, the Hawking-Gibbons temperature of the semiclassical gravity QFT regime
becomes the intrinsic string temperature of the quantum string regime. \\
Also, in the AdS background, the ${\cal R}$-transformation manifests as well between $T_H$ and $T_S$. \\ \\
When back reaction is included, the $\mathcal{R}$-dual relations between 
semiclassical (QFT) and stringy phases manifest as well. \\ \\
These BH and dS examples suggest that our ${\cal R}$ transform could be promoted to a dynamical
operation: evolution from a semiclassical gravity - QFT phase to a quantum gravity string phase (as in BH 
evaporation) or conversely, evolution from a quantum string phase to a semiclassical QFT phase (inflation) 
(as in cosmological evolution). These issues, as well as other unifying concepts between black holes 
and elementary particles, have been further developped and elaborated in Ref {17}, including the 
corresponding density of states and  entropies in the both regimes, and the quantum dS and AdS string states. 
The last BH string state decays with a 
decay rate which is precisely the ${\cal R}$ transformed of the semiclassical thermal BH decay formula, 
\\

A particular consequence of these results is that there is no lost of information in black hole evaporation (there is no 
paradox at all).The results by these authors on the last stage of black hole evaporation,  its emission spectrum and its 
final decay in pure radiation, are  reported in Ref [4] and in Ref [17]. Recently, Stephen 
Hawking, Ref [18] presented, within another approach  (euclidean QFT gravity), his resolution of the 
``information paradox'', a problem which himself have clearly posed and pioneered. These are very important news for the 
subject, 
and encouradge new roads to the understanding of quantum black holes, very early cosmology and the classical-quantum dual
nature of Nature .

\vskip 2cm

{\bf References}

\bigskip \bigskip

1. N.D. Birrell and P.C.W. Davies, {\it Quantum Fields in Curved Space}, (Cambridge
   University Press, England, 1982).
\medskip

2. {\it String Theory in Curved Space Times}, edited by N. S\'anchez (World Scientific. Pub, 
Singapore, 1998).
\medskip

3. M. Ramon Medrano and N. S\'anchez, Phys. Rev.  D60, 125014 (1999).
\medskip

4. M. Ramon Medrano and N. S\'anchez, Phys. Rev. D61, 084030 (2000).
\medskip

5. H.J. de Vega and N. S\'anchez, Nucl. Phys. B309, 522 (1988); B309, 577 (1988).
\medskip

6. A.L. Larsen and N. S\'anchez, Phys. Rev. D52, 1051 (1995).
\medskip

7. H.J. de Vega, A.L. Larsen and N. S\'anchez, Phys. Rev. D15, 6917 (1995).
\medskip

8. H.J. de Vega and N. S\'anchez, Phys. Lett. B197, 320 (1987).
\medskip

9. H.J. de Vega, L. Larsen, N. S\'anchez, Phys Rev. D58, 026001(1998).
\medskip

10. S. W. Hawking, Comm. Math. Phys. 43, 199 (1975).
\medskip

11. G.W. Gibbons and S. W. Hawking, Phys. Rev. D15, 2738 (1977).
\medskip

12. T.H. Buscher, Phys Lett B159, 127 (1985); B201, 466 (1988), B194, 51 (1987).
\medskip

13. See for example J. Polchinski, {\it String Theory} (Vol I and II. Cambridge University 
    Press, 1998), and references therein.
\medskip

14. J. Maldacena, Adv Theor Math Phys 2, 231 (1998). 
\medskip

15.  See for example E. Papantonopoulos, V. Zamarias, hep-th/0307144 
     and references therein. 
\medskip

16. M. Ramon Medrano and N.G. S\'anchez, Mod. Phys. Lett A18, 2537 (2003).
\medskip

17. N.G. S\'anchez, {\it``Conceptual unification of elementary particles, black holes}, 
   {\it quantum de Sitter and Anti-de Sitter string states''} hep-th/0312018.

\medskip 
18. S.W. Hawking, communication , GR17 Conference, Dublin, 18-24 July (2004).
\medskip
\end{document}